\documentstyle[12pt]{article}
\def\bc{\begin{center}}
\def\ec{\end{center}}
\def\be{\begin{eqnarray}}
\def\ee{\end{eqnarray}}

\def\d#1#2{\frac{\displaystyle #1}{\displaystyle #2}}

\def\diff{di\hspace{-0.2em}f\hspace{-0.3em}f}

\def\r{\partial}

\def\Si{\Sigma}
\def\Om{\Omega}

\def\dl{\delta}
\def\eps{\epsilon}

\def\del{\nabla}

\def\PRD{{\it Phys. Rev.}~{\bf D}}

\def\CQG{{\it Class. Quantum Gravity }}

\begin{document}
\hfill 
August 22, 2002
\bc{\Large \bf Noether-Charge Realization of
\\Diffeomorphism 
Algebra 
} \ec

\bigskip
\begin{center}
Han-Ying Guo$^b$\footnote{Email:
hyguo@itp.ac.cn}, \quad Chao-Guang Huang$^{a,
b}$\footnote{Email: huangcg@mail.ihep.ac.cn}\quad
and
 \quad Xiaoning Wu$^{c}$\footnote{Email:
wuxiaon@yahoo.com.cn}
\end{center}
\begin{center}{\small
a. Institute of High Energy Physics, Chinese
Academy of
Sciences,\\
\quad P.O. Box 918(4), Beijing 100039, China.\\
b. Institute of Theoretical Physics, Chinese Academy of Sciences,\\
\quad  P.O. Box 2735, Beijing 100080,
China.\\
c. Max-Plank-Institut f\"ur Gravitationsphysik,
Am M\"uhlenberg 1, 14476 Golm, Germany\\
}
\end{center}
\hspace{2cm}

\bigskip\bigskip

\bc
\parbox{13cm}
{{\bf Abstract.}\quad  It is shown in the covariant phase space
formalism that the Noether charges with respect to the diffeomorphism
generated by vector fields and their horizontal variations in general
relativity form a diffeomorphism algebra.
It is also shown with the help of the null tetrad which is well
defined everywhere that the central term of the reduced
diffeomorphism algebra on the Killing horizon for a large class of
vector fields vanishes.
} \ec

\bigskip
\bc
\parbox{13cm}
\ec



\section{Introduction}

\quad ~ It is 
well known that in the
canonical formalism of general relativity (GR),
the Hamiltonian constraint ${\cal H} \approx 0$
and 3-dimensional diffeomorphism constraint
${\cal H}_i \approx 0$ form an algebra under the
Poisson bracket \cite{{adm},{T}}, which reflects
the diffeomorphism structure of space-time.
However, the canonical approach does not preserve
the covariance of GR manifestly.  To recover the
manifest covariance of the theory, a new
approach, the covariant phase space formalism,
has been developed \cite{CE}--\cite{iw2}.

In studying the statistical
origin of the black hole entropy 
by use of the covariant phase space method 
\cite{c2}, 
Carlip treated the Hamiltonian functional
conjugate to a vector field 
the
generator of the diffeomorphism algebra just like
the Hamiltonian in the canonical approach. The
diffeomorphism algebra is assumed to be realized
by the Poisson bracket or by the Dirac bracket
\cite{{Db},{BHS}} on the constraint surface.
Unfortunately, the Hamiltonian functional
conjugate to a vector field $\xi^a$ does not
always exist, as pointed out by Wald \cite{wald},
for a given boundary condition.  In addition, in
the lack of the definition of the Poisson bracket
and thus the Dirac bracket in the covariant phase
space formalism, Carlip borrowed the Poisson
bracket and the Dirac bracket from the ADM
formalism \cite{{adm},{T}}.  

On the other hand, in both classical and quantum
field theories, if a Lagrangian possesses certain
symmetries, such as gauge symmetry and
Poincar\'{e} symmetry, the corresponding Lie
algebras can always be generated by the Noether
charges of the conservation currents with respect
to the symmetries. This spirit should also be
available for the diffeomorphism invariance of a
diffeomorphism invariant theory such as GR since
the set of diffeomorphisms forms an infinite
dimensional group under composition \cite{amr}.
The main purpose of the present letter is to show
how the algebra $\diff$(${\cal M}$)  can be
realized from the Noether charges in GR.
It is also shown with the help of
the null tetrad which is well defined everywhere
that the central term of the reduced
diffeomorphism algebra on the Killing horizon for
a large class of
vector fields vanishes.

The letter is arranged in the following way. In the next section,
we briefly review the Noether currents with respect to the
diffeomorphisms generated by vector fields and their charges in a
diffeomorphism invariant theory.  In section 3, we show that the
diffeomorphism algebra may be realized by the Noether charges and
their horizontal variations.  In section 4, we 
study the reduced algebra on the event horizon of black-hole
space-time manifold.
In the last section, some remarks are given.

\section{Noether Currents and Their Charges}

\quad ~ Let ${\bf L}$ be the Lagrangian 4-form of
a diffeomorphism-invariant gravitational metric
theory. Its horizontal variation induced by the
vector
field $\xi^a$ can be written as 
\cite{{wald},{iw},{wghw}}
\begin{eqnarray}
\label{f1} \hat \dl_\xi {\bf L}={\bf E} \hat
\dl_\xi g+ d{\mbox {\boldmath $\Theta$}}(g,
\hat\dl_\xi g), \ee where ${\bf E}=0$ gives rise
to the Euler-Lagrange equation for the
theory and ${\mbox {\boldmath
$\Theta$}}(g,\hat \dl_\xi g)$ is the symplectic
potential 3-form. On the other hand, using the
Lie derivative ${\cal L}_\xi$  \be \label{f2}
\hat\dl_\xi{\bf L} ={\cal L}_\xi{\bf L}= d(\xi
\cdot {\bf L}).
\end{eqnarray}
 Equating Eqs.(\ref{f1})
and (\ref{f2}), one gets
\begin{eqnarray}
\label{Neq}
d*{\bf j}(\xi) +{\bf E}\hat\dl_\xi g=0,
\end{eqnarray}
where \be \label{cc1} {\bf j}(\xi) = *({\mbox
{\boldmath $\Theta$}}(g,\hat \dl_\xi g) -\xi
\cdot {\bf L}) \ee is the Noether current 1-form
with respect to the diffeomorphism generated by a
given vector field $\xi^a$. Its (entire) Noether
charge is given by the integral over a Cauchy
surface $\Si$ \be \label{NC1}
Q(\xi)=\int_{\Sigma} *{\bf j}(\xi). \ee

In vacuum GR, the Lagrangian 4-form in units of
$G=c=1$ reads
\be
{\bf L} = \d 1 {16\pi} R {\mbox
{\boldmath $\eps$}},
\ee
where $R$ is the scalar
curvature and ${\mbox {\boldmath $\eps$}}$  the
volume 4-form.  The symplectic potential takes the
form of
\be
{\mbox {\boldmath $\Theta$}}_{abc}
(g, {\cal L}_{\xi}g) = \d 1 {16\pi}[\del^d
(g_{ef} \hat \dl_\xi g^{ef})-\del_e \hat \dl_\xi
g^{de}] {\mbox {\boldmath $\eps$}}_{dabc}.
\ee
Thus, the Noether current and the Noether charge may be
explicitly written as
\be
\label{cc2}
{\bf j}_{a}(\xi) =
\frac{1}{8\pi}G_{ab}\xi^b+\frac{1}{16\pi}
(\nabla^b\nabla_a\xi_b-\nabla^b\nabla_b\xi_a)
\ee
and
\be
\label{NC2}
Q(\xi) = \d 1 {8\pi}\int _\Si * (G_{ab} \xi^b)
- \frac{1}{16 \pi}\int _{\r \Si} *d \xi,
\ee
respectively, where $G_{ab}$ is the Einstein
tensor and $\r \Si$  the boundary of the Cauchy
surface. On shell, the first terms in Eqs. (\ref{cc2})
and (\ref{NC2}) vanish and the Noether charge may
be expressed as an integral over the boundary of
the Cauchy surface.
The Noether current (\ref{cc2}) is exactly the
dual of the current 3-form constructed by Wald
{\it et al} earlier (see, for example,
\cite{{wald}, {iw}}).  The integrand of the second
term in
(\ref{NC2}) is nothing else but the Noether
charge 2-form in \cite{{wald}, {iw}}. The
boundary $\r \Si$, in general, consists of two
closed 2-dimensional surfaces at the two ends of
the Cauchy surface. 
For the whole
asymptotically flat region, the Cauchy surface
emanates from the bifurcation surface and extends
to the spatial infinity.  Thus, the boundary of
the Cauchy surface $\partial \Sigma$ should be
$S_H^{(-)}\cup S_\infty$, where $S_H^{(-)}$ is
the bifurcation surface.  The superscript $(-)$
stands for the opposite orientation. Namely, its
normal vector points to the direction of $r$
decreasing.  Then, the Noether charge becomes
the algebraic summation of the partial Noether
charges of the closed surfaces, i.e.
\be
\label{NC2o}
Q(\xi) = {\cal Q}_{\infty}(\xi)-{\cal Q}_{H}(\xi).
\ee

In particular, for the stationary, axisymmetric
black hole space-time with Killing vector \be
\label{kv}
\chi_K ^a = t_K^a + \Om_H \phi_K^a,
\ee
where $t_K^a$ and $\phi_K^a$ are the time-like and
space-like Killing vector of the space-time,
respectively, $\Om_H$  the angular velocity on
the horizon, \be
\frac{1}{2}(\nabla_b\nabla^a\chi_K^b-\nabla_b\nabla^b\chi_K^a)
= R^a_b\chi_K^b. \ee From Eq. (\ref{cc2}), it
follows that ${\bf j_a}$ and thus $Q(\chi_K)$
vanishes on shell.  Since the mass and the
angular momentum of the black hole are, by
definition,
\begin{eqnarray}
M=- \frac{1}{8\pi}\int_{S_{\infty}}*dt_K
\ee
and
\be
J=\frac{1}{16\pi}\int_{S_{\infty}}*d\phi_K,
\end{eqnarray}
respectively, the expression for ${\cal Q}_\infty$ is then
\begin{eqnarray}
{\cal Q}_\infty=-\frac{1}{16\pi}\int_{S_{\infty}}*d\chi_K
 =\frac 1 2 M - \Omega_HJ.
\end{eqnarray}
On the other hand, the expression for ${\cal Q}_H$ takes the
form of
\begin{eqnarray}
{\cal Q}_H = - \frac{1}{16\pi}\int_{S_H}*d\chi_K
=\frac{\kappa}{8\pi}A,
\end{eqnarray}
where $\kappa$ is the surface gravity and $A$ the
area of the cross section of the event horizon.
Thus, the vanishing Noether charge $Q(\chi_K) =
0$ gives rise to the mass formula \cite{wghw}
\begin{equation}\label{mass}
Q= \frac 1 2 M- \Omega_HJ-\frac{\kappa}{8\pi}A=0.
\end{equation}

The Noether charge may be defined for a finite
region of a space-time.  When $\Sigma$ is not
chosen to be the whole of the Cauchy surface but
a partial Cauchy surface $\Si_p$ with two
boundaries $B_1$ and $B_2$ such that
$A_{B_2}>A_{B_1}$, where $A_{B_i}$ stands for the
area of surface $B_i$, \be \label{NC3}
Q_{\Si_p}(\xi) = {\cal Q}_{B_2}(\xi) - {\cal
Q}_{B_1}(\xi) \ee gives the Noether charge for
the portion of the space-time region
$R\times\Si_p$.

\section{Realization of Algebra $\diff$(${\cal M}$)}

\quad ~ In order to get the realization of
algebra $\diff$(${\cal M}$), it is needed to
consider the two successive horizontal variations
of the Lagrangian 4-form induced by two vector
fields $\xi_1$ and $\xi_2$. 
Due to the property
\begin{eqnarray} \label{Ldr}
[\hat \dl_{\xi_1}, \hat \dl_{\xi_2}]{\bf L} =
\hat \dl_{[\xi_1, \xi_2]}{\bf L},
\end{eqnarray}
it is straightforward to get 
\begin{eqnarray}
\label{dEdg}
d\{\hat \dl_{\xi_1}[*{\bf j}(\xi_2)]-
\hat \dl_{\xi_2}[*{\bf j}(\xi_1)]
-*{\bf j}([\xi_1,\xi_2])\} = \hat \dl_{\xi_2}({\bf
E}\hat \dl_{\xi_1} g) - \hat \dl_{\xi_1}({\bf
E}\hat \dl_{\xi_2} g)+{\bf E}\hat \dl_{[\xi_1,
\xi_2]} g.
\end{eqnarray}
Namely, the  combination of the current 1-forms
\be
* \hat \dl_{\xi_1} [*{\bf j}(\xi_2)] -
* \hat \dl_{\xi_1} [*{\bf j}(\xi_2)] -
{\bf j}([\xi_1,\xi_2])
\ee
is also conserved on shell.  The Noether-like
charge
$-K(\xi_1,\xi_2)$ with respect to this combination 
on a Cauchy surface $\Sigma$ is then given by
\be
\label{ac0}
- K(\xi_1, \xi_2) = \hat \dl _{\xi_1}
Q(\xi_2) - \hat \dl _{\xi_2} Q(\xi_1) - Q([\xi_1,
\xi_2]). \ee It should be noted that in
Eq.(\ref{ac0}) the (horizontal) variation $\hat
\dl_{\xi_1}$ acts on both $Q$ and $\xi_2$.
Namely, \be \hat \dl _{\xi_1} Q(\xi_2) = (\hat
\dl _{\xi_1} Q)(\xi_2) + Q(\hat \dl _{\xi_1}
\xi_2) =(\hat \dl _{\xi_1} Q)(\xi_2)+
Q([\xi_1,\xi_2]). \ee Hence, Eq.(\ref{ac0}) gives
rise to \be \label{ac} (\hat \dl _{\xi_2}
Q)(\xi_1) - (\hat \dl _{\xi_1} Q)(\xi_2)
=Q([\xi_1,\xi_2])+K(\xi_1,\xi_2).
\ee
Eq.(\ref{ac}) shows that the Noether charges and
their horizontal variations form an algebraic
relation of $\diff$(${\cal M}$) and that
$K(\xi_1, \xi_2)$ may be treated as the possible
central term.   (Note that in Eq.(\ref{ac}) the
vector fields keep unchanged under the variation,
i.e. $\hat \dl _{\xi_1} \xi_2=\hat \dl _{\xi_2}
\xi_1=0$ \cite{wald}.)

Further, the Jacobi identity of the horizontal variations
\begin{eqnarray}
\label{jcb}
([\hat\delta_{\xi_1},[\hat\delta_{\xi_2},
\hat\delta_{\xi_3}]]+[\hat\delta_{\xi_2},[\hat\delta_{\xi_3},
\hat\delta_{\xi_1}]]+[\hat\delta_{\xi_3},[\hat\delta_{\xi_1},
\hat\delta_{\xi_2}]]) {\bf L}=0
\end{eqnarray}
results in
\begin{eqnarray}\label{cce}
K([\xi_1,\xi_2], \xi_3)+K([\xi_2,\xi_3],
\xi_1)+K([\xi_3,\xi_1], \xi_2)=0.
\end{eqnarray}
This is the two co-cycle condition for the center
term.

To determine the possible central term, one has to calculate
$(\hat \dl _{\xi_2} Q)(\xi_1) - (\hat \dl _{\xi_1} Q)(\xi_2)$ and
$Q([\xi_1,\xi_2])$.   By definition and the conservation equation
$d*{\bf j}=0$,
\be
\label{dlQ}
(\hat \dl _{\xi_2} Q)(\xi_1) - (\hat \dl _{\xi_1} Q)(\xi_2) =
\int _{\r \Si} \xi_2 \cdot
{\mbox {\boldmath $\Theta$}} (g, {\cal L}_{\xi_1}g) -
\xi_1 \cdot {\mbox {\boldmath $\Theta$}}
(g, {\cal L}_{\xi_2}g) + \xi_1 \cdot (\xi_2 \cdot {\bf L}) -
\xi_2 \cdot (\xi_1 \cdot {\bf L}).
\ee

For vacuum GR, the Lagrangian
${\bf L}$ vanishes on-shell and thus
\be
\label{Qb0}
(\hat \dl _{\xi_2} Q)(\xi_1) - (\hat \dl _{\xi_1} Q)(\xi_2) =
\d 1 {16 \pi} \int_{\r \Si} {\mbox {\boldmath $\eps$}}_{abcd}
[\xi^c_2\del_e(\del^e\xi^d_1-\del^d\xi^e_1)
- \xi^c_1\del_e(\del^e\xi^d_2-\del^d\xi^e_2)].
\ee
On the other hand,
\be
Q([\xi_1,\xi_2])= -\frac{1}{16 \pi}\int _{\r \Si}
{\mbox {\boldmath $\epsilon$}} _{abcd}\del^c
(\xi_1^e \del_e \xi_2^d - \xi_2^e \del_e \xi_1^d ).
\ee
The possible central term is then
\be
\label{Qa}
K(\xi_1,\xi_2) &=& (\hat \dl _{\xi_2} Q)(\xi_1) -
(\hat \dl _{\xi_1} Q)(\xi_2) - Q([\xi_1,\xi_2])
\nonumber \\
&=& \frac{1}{8 \pi}\int _{\r \Si}
{\mbox {\boldmath $\epsilon$}} _{abcd} \left [
\del _e (\xi_{[1}^c \del ^d \xi_{2]}^e) -
2 \xi_{[1}^{[c} \del ^{e]} \del_{|e|} \xi_{2]}^{d}
\right ].
\ee

The above analysis may also apply to a hyperbolic region
with a partial Cauchy surface $\Si_p$ in the manifold.  Namely,
the Noether charges $Q_{\Sigma_p}(\xi)$s for the hyperbolic
region with a partial Cauchy surface and their horizontal
variation form the algebraic relation of $\diff$($R\times\Si_p$),
\be
\label{Qb1}
(\hat \dl _{\xi_2}Q_{\Sigma_p})(\xi_1)-(\hat
\dl_{\xi_1}Q_{\Sigma_p})(\xi_2)=
Q_{\Sigma_p}([\xi_1,\xi_2]) + K_{\Si_p}(\xi_1,\xi_2).
\ee
Since $Q_{\Si_p}$ is expressed on shell in terms
of the algebraic summation of the boundary terms as Eq.
(\ref{NC3}), Eq. (\ref{Qb1}) can be separated into two
algebraic relations
\be
\label{Qv3}
(\hat \dl _{\xi_2} {\cal Q}_{B_i})(\xi_1) -
(\hat \dl _{\xi_1} {\cal Q}_{B_i})(\xi_2)
={\cal Q}_{B_i}([\xi_1,\xi_2]) +
K_{B_i}(\xi_1,\xi_2), \qquad i=1,2
\ee
with help of Eq. (\ref{NC3}).

\section{Reduced algebra on event horizon of stationary
axisymmetric black hole}

\quad ~ Let us consider such a partial Cauchy
surface $\Si_1$ that it emanates from the
bifurcation surface $S_H$, extends almost along
the generator of event horizon of a black hole
and ends at certain place of the stretched
Killing horizon \cite{c2} denoted by
$B_\epsilon$. At the end of calculation, this
partial Cauchy surface tends to the event horizon
by taking $\epsilon \rightarrow 0$.
Define a vector field orthogonal to the
Killing vector fields (\ref{kv}) by
\begin{eqnarray} \label{Rn}
\del_a \chi_K^2 = -2 \kappa \rho_a,
\end{eqnarray}
where $\kappa$ is the surface gravity on the event horizon.
The two vector fields $\chi_K^a$ and $\rho^a$ may combine
into the two null vectors,
\be
l^a &=& \d 1 2 (\chi_K^a + \d {|\chi_K|}{\rho} \rho^a) \nonumber \\
n^a &=& -\d 1 {\chi_K^2}(\chi_K^a -\d {|\chi_K|}{\rho}\rho^a).
\ee
$l^a, n^a$ and $m^a, \bar m^a$ constitue a null tetrad field in the
neigborhood of the event horizon.  The Lie bracket of $l^a$ and
$n^a$ reads
\be
\ [l, n]^a = - \kappa \d {\rho}{|\chi|} n^a.
\ee
It may also be checked that
\be
D\chi_K^2 & := &l^a \nabla_a \chi_K^2  = O(\chi_K^2) \nonumber \\
\Delta \chi_K^2 & := &n^a \nabla_a \chi_K^2 =  O(1).
\ee
Therefore, the vector fields of type
\be
\xi^a = T l^a + R  n^a    \qquad  \mbox{with $R\sim O(\chi_K^2)$}
\ee
form a closed algebra under the Lie bracket.

For this type of vector fields,
the partial Noether charge ${\cal Q}$ reads
\be
{\cal Q}_S(\xi) =-\d 1 {16\pi} \int_S {\mbox {\boldmath $\hat \eps$}}_{ab}
(DT-\Delta R + \kappa T)
\ee
and Eq. (\ref{Qb0}) for the partial Noether charge ${\cal Q}$
reduces to
\begin{eqnarray} \label{Qb5}
(\hat \dl _{\xi_2} {\cal
Q}_S)(\xi_1) - (\hat \dl _{\xi_1} {\cal Q}_S)(\xi_2)
 = \d 1 {8\pi}
\int_S {\mbox {\boldmath $\hat \eps$}}_{ab} (T_{[1} D \Delta R_{2]} -T_{[1} D^2 T_{2]} -
\kappa T_{[1} DT_{2]})
\end{eqnarray}
when the boundary condition $l_{a;\ b}(m^a\bar m^b+\bar m^a m^b)|_{S}=0$ is satisfied,
where 
$S=S_H$ or ${\displaystyle \lim_{\eps \to 0} {B_\epsilon}}$
(denoted by $B$ hereafter), and
${\mbox {\boldmath $\hat \epsilon$}}_{ab}$ is the area 2-form of
$S$.
The straightforward calculation shows
\begin{eqnarray} \label{nc}
{\cal Q}_S([\xi_1,\xi_2])= - \d 1 {8\pi}
\int_S {\mbox {\boldmath $\hat \eps$}}_{ab} (-T_{[1} D \Delta R_{2]} +T_{[1} D^2 T_{2]} +
\kappa T_{[1} DT_{2]}).
\end{eqnarray}
Thus, the central term vanishes!

\section{Concluding Remarks}

\quad ~ In conclusion, as other symmetries in other classical and
quantum field theories, the diffeomorphism algebra, reflecting
the diffeomorphism invariance of diffeomorphism invariant theories,
may be realized by the Noether charges and their horizontal variations.
The Noether-charge realization has four
remarkable features. First of all, the
Noether-charge realization always exists because
the Noether charge always exists for any given
vector fields and any given boundary conditions.
This is in contrast to the Hamiltonian-functional
realization, which does not exist for some vector
fields and boundary conditions because the
Hamiltonian functionals themselves do not always
exist \cite{wald}.  Secondly, the Noether-charge
realization is a completely covariant approach.
In the present approach, the Poisson bracket and
Dirac bracket are not used at all, which
are defined in the
canonical approach. 
Thirdly, only the horizontal variations are
considered in the Noether-charge realization.
Finally, for vacuum general relativity the
Noether-charge realization has the same form as
the one given by Carlip with the help of the
Hamiltonian functionals \cite{c2}.

Another conclusion of the present letter is that
the central term on Killing horizon for a large
class of vector fields vanishes! The key point is
that the null tetrad instead of the basis $\{
\chi _K^a, \rho ^a, t^a_1, t_2^a \}$ is used. The
former is well defined everywhere, including on
the Killing horizon, while the latter is
ill-defined on the Killing horizon. Therefore,
the appearance of the central term seems to come
from the choice of basis.

\section*{Acknowledgments}

\quad   We would like to thank Professors/Drs. Z. Chang,
C.-B. Liang,  Y. Ling, R.-S. Tung,  K. Wu, H.-X. Yang,
M. Yu, B. Zhou, and C.-J. Zhu for helpful discussion.
This project is in part supported by National Science
Foundation of China under Grants Nos. 90103004,
10175070.

\end{document}